\documentclass[aps,prl,twocolumn,superscriptaddress,showpacs]{revtex4}
\usepackage{graphicx}
\usepackage{amsmath}
\usepackage{amsfonts}
\usepackage{amssymb}
\usepackage{bm}
\usepackage{color}

\newcommand\ii{{\mathrm{i}}}

\begin{document}

\title{Fluctuation spectroscopy of granularity in superconducting structures}
\author{I.~V.~Lerner}
\affiliation{School of Physics and Astronomy, University of Birmingham, Edgbaston, B15
2TT Birmingham, UK}
\affiliation{Materials Science Division, Argonne National Laboratory, 9700 S.Cass Avenue,
Argonne Il 60439}
\author{A.~A.~Varlamov}
\affiliation{COHERENTIA-INFM, CNR, Viale del Politecnico 1, I-00133 Rome, Italy}
\affiliation{Materials Science Division, Argonne National Laboratory, 9700 S.Cass Avenue,
Argonne Il 60439}
\author{V.~M.~Vinokur}
\affiliation{Materials Science Division, Argonne National Laboratory, 9700 S.Cass Avenue,
Argonne Il 60439}
\date{submitted 16 July 2007; published 21 March 2008}

\begin{abstract}
We suggest to use `fluctuation spectroscopy' as a method to detect
granularity in a disordered metal close to a superconducting transition.
We show that with lowering temperature $T$  the resistance $R(T)$  of a system of
relatively large grains initially grows due to the fluctuation suppression
of the one-electron tunneling but decreases with
further lowering $T$  due to the coherent charge transfer of the fluctuation Cooper pairs.  Under certain conditions, such a maximum in $R(T)$ turns out to be sensitive to weak magnetic fields due to a novel Maki -- Thompson type mechanism.
\end{abstract}

\maketitle

Since seminal experiments on the superconducting-insulator transition in
granular samples~\cite{gold}, transport properties of granular metals enjoy
extensive attention, see \cite{revG,rev} for reviews and references.
Granularity offers systems with tuneable parameters eminently suitable for
studies of the interplay between electron correlations and mesoscopic
effects of disorder. Generically, effects of granularity are most profound
at $T\gtrsim \Gamma$ ($\Gamma$ is the electron tunneling rate between the
grains). At low temperatures ($T\lesssim \Gamma$) transport properties of
the granular metal coincide with those of amorphous one \cite{BELV:03}
including rather subtle effects of weak localization \cite{Biagini,Vinokur2}%
. The role of granularity is widely believed to be crucial for the
metal-insulator and superconductor-insulator transitions. It was recently
discovered \cite{bat,gold07} that it may be of self-induced nature and appear even
in homogeneously disordered samples. This poses an interesting question
whether it is possible to find a reliable experimental benchmark of
granularity in electronic transport in the metallic regime.

In this Letter we show that there exists a firm experimental signature of
granularity -- the appearance of a maximum in the temperature dependence of the resistance $R(T)$, provided that a granular system is made of material which experiences the
superconducting transition. The magnitude and position of such a  maximum might be very sensitive to a weak magnetic field. The maximum is due to an interplay of different types of
superconducting fluctuation contributions,  specific for granular systems,   at temperatures $T\gtrsim T_{\text{c}} $  (the transition temperature). All this is strikingly different from a monotonic decrease with $T$ of the fluctuation resistance of amorphous systems close to $T_{\text{c}} $ and allows one to extract various characteristics of the granularity, thus suggesting a method of characterization of disordered systems which can be termed as the fluctuation spectroscopy.

The disordered system becomes effectively granular for one-electron transport provided that the tunneling conductance between inhomogeneities (grains), $g_{\text{T}}\sim \Gamma/\delta$, is much smaller than the intragranular conductance $g\sim E_{\text{Th}}/\delta $ -- in the opposite case, the system is indistinguishable from the amorphous one. Furthermore, the granular character is preserved also for the charge transfer by fluctuation  Cooper pairs when  their Ginzburg -- Landau life-time, $\tau _{\text{GL}}  \sim (T-T_{\text{c}})^{-1}  $ is much shorter than the escape time $\Gamma^{-1} $ (we use units with $\hbar=k_{\text{B}} =1$). Finally, the system remains metallic when $g_{\text{T}}\gg1 $. The above conditions can be satisfied when
\begin{align}\label{0}
    \delta\lesssim \Gamma \lesssim E_{\text{Th}}, T_{\text{c}}\,.
\end{align}
Here $\delta$ is the mean level spacing in the grain, $E_{\text{Th}}=\mathcal{D}/d^2 $ is its Thouless energy,  $d$ is
the typical grain size, and $\mathcal{D}$
is the intragrain diffusion coefficient \cite{noteLVV1}. We will focus at the case $E_{\text{Th}}\lesssim T_{\text{c}}  $ where the Cooper pair intragrain motion can have both the  three-dimensional ($3D$) and zero-dimensional ($0D$) character.

The  characteristic feature of the fluctuation pairing in the problem is the appearance of two different scales
for the superconducting coherence length, $\xi_{ \text{g}}=(\mathcal{D}/T_{\text{c}})^{1/2}  $ and $\xi_{\text{T
}} =(\Gamma d^2/T_{\text{c}} )^{1/2} $, driven by the intragrain and intergrain pairing, respectively.   The above inequalities correspond to the length scales ranging as $\xi_{\text{T}} \lesssim \xi_{\text{g}}\lesssim d$.

The existence of the two correlation scales becomes crucial in the vicinity of $T_{\text{c}}$ where
the superconducting fluctuations are governed by the temperature-dependent
Ginzburg-Landau correlation length $\xi (\epsilon )$:
\begin{align*}
\xi _{\text{g}}(\epsilon )&\equiv \frac{\xi _{\text{g}}}{\sqrt{\epsilon }}\,,
& \xi _{\text{T}}(\epsilon )&\equiv \frac{\xi _{\text{T}}}{\sqrt{\epsilon }}
\,, & \epsilon &\equiv \frac{T-T_{\text{c}}}{T_{\text{c}}}\,.
\end{align*}
This leads to the existence of three distinct temperature regimes near $T_{
\text{c}}$ (see Fig.~1):
\begin{align}  \epsilon _{\text{g}}\lesssim \epsilon& \lesssim
1\,, &&(3D) \label{1}
    \\
    \epsilon _{\text{T}
}\lesssim \epsilon &\lesssim \epsilon _{\text{g}} \,, &&(0D)\label{2}\\
\epsilon &\lesssim \epsilon _{\text{T}}\,, &&(3D)\label{3}
\end{align}
where $\epsilon_{\text{T}}\equiv \Gamma/T_{\text{c}} $ and $\epsilon_{\text{g}}\equiv E_{\text{Th}}/T_{\text{c}}  $.
For the first two regimes, it is the intragrain
correlations  which govern the
fluctuation corrections to the conductance,
these correlations having a $3D$  character in region (\ref{1}), albeit very different from that in the amorphous metal, and a $0D$ character in region (\ref{2}). For smaller grains, when $E_{\text{Th}}\gtrsim T_{\text{c}}  $, region (\ref{1}) vanishes which does not change the qualitative picture.
In the immediate vicinity of $T_{\text{c}}$, regime (\ref{3}), the intergrain fluctuation pairing dominates. In this case, the fluctuation corrections are equivalent to those in amorphous disordered media with the effective diffusion coefficient $D_{\text{T}}=\Gamma d^2$. It is  the existence of the regimes (\ref{1}) and (\ref{2}) which imprints a pronounced signature of the granularity as we now demonstrate.

\begin{figure}[t]\vspace*{1mm}

 \includegraphics[width=.6\columnwidth,angle=270]{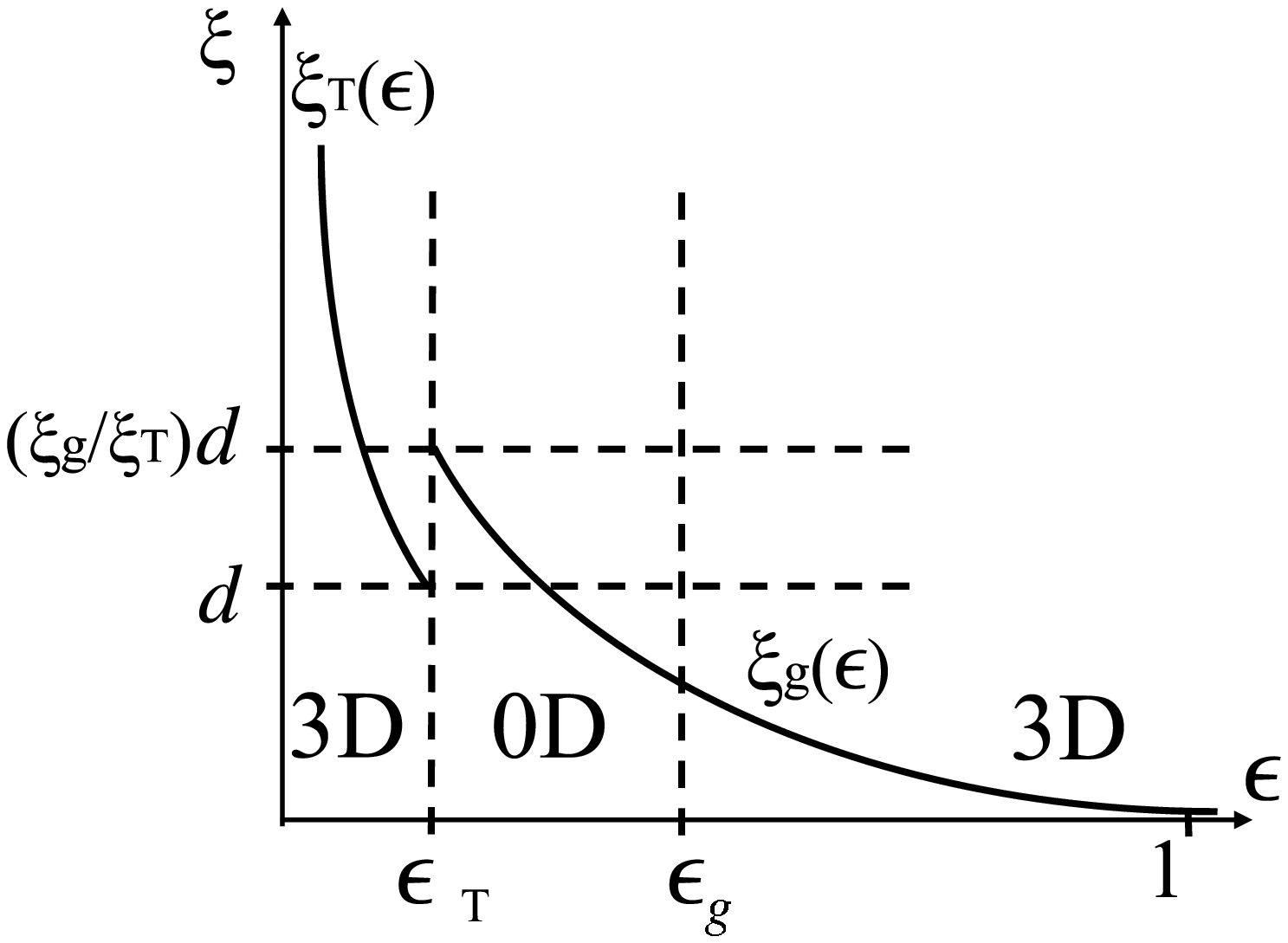}
\caption{Temperature dependence of the  Ginzburg--Landau superconducting
coherence length above $T_{\text{c}} $: at $\epsilon=\epsilon_{\text{T}} $ it jumps from the intra- to inter-granular regime.
 }
\label{cohl}
\end{figure}

The model under consideration is defined by the standard Hamiltonian,
\begin{align*}
\hat{H}=\hat{H}_0+\hat{H}_{\text{T}}\,.
\end{align*}
where $\hat{H}_0$ describes electrons in a single grain in the presence of
the BCS pairing (characterized by the transition temperature $T_{\text{c}}$)
and disorder (characterized by the intragrain diffusion coefficient $%
\mathcal{D}$), while $\hat{H}_{\text{T}}$ describes the intergrain tunneling
with the amplitude $t$ which is assumed to be local (\emph{i.e.}\
momentum-independent) and the same for all the grain pairs. For simplicity,
the grains are assumed to have a spherical shape of diameter $d$ \cite%
{noteLVV2}.

The tunneling current between the neighboring grains can be expressed as
follows \cite{VD83}:
\begin{equation*}
I\left( V\right) =-e{\operatorname{Im}}K^{\text{R}}\left( \omega \right) \Bigr|%
_{\omega =-{\mathrm{i}}{}eV}\,,
\end{equation*}%
where $K^{\text{R}}(\omega )$ is the analytical continuation, $\ii \omega_\nu\to\omega$,  to the upper
half-plane of the current-current correlation function in the Matsubara
frequencies
\begin{equation}\label{K}
\mathcal{K}\left( \omega _{\nu}\right) =T|t|^{2}\sum_{\varepsilon _{n}}\sum_{%
\bm {p,p'}}\mathcal{G}_{L}\left( \bm {p},\varepsilon _{n}+\omega _{\nu}\right)
\mathcal{G}_{R}\left( \bm {p'},\varepsilon _{n}\right) \,.
\end{equation}%
Here $\mathcal{G}_{L,R}$ are the exact electron Green's functions in the
neighboring (``left'' and ``right'') grains with $\varepsilon _{n}=\pi T(2n+1)$ and $%
\omega _{m}=2\pi Tm$ being the fermionic and bosonic Matsubara frequencies.
Hence the conductivity
\begin{equation}\notag
\sigma _{\text{T}}=\frac{1}{d}\frac{{\operatorname{d}}I}{{\operatorname{d}}V}=\frac{e^{2}}{d}%
\left. {\operatorname{Im}}\frac{{\operatorname{d}}K^{R}\left( \omega \right) }{{}{\operatorname{d}}%
\omega }\right\vert _{\omega \rightarrow 0}\,, % \label{G}
\end{equation}%
where $V$ is the voltage drop at the grain.

\begin{figure}[b]
\includegraphics[width=.9\columnwidth]{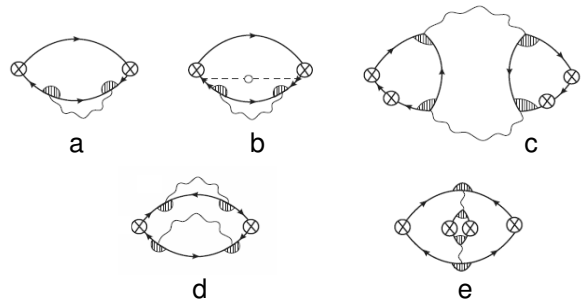}
\caption{Fluctuation corrections to the tunneling current  in the lowest orders in transparency and fluctuations.}
\label{tundia}
\end{figure}

Examples of diagrams describing significant fluctuation contributions to $%
\mathcal{K}\left( \omega _{m}\right) $ are presented in Fig.~\ref{tundia}.
The solid lines correspond to the disorder-averaged electron Green's
functions, the circled crosses represent the tunneling amplitude $t$, and
the wavy lines correspond to the fluctuation propagator, $\mathcal{L}\left( q_{k},\Omega_n \right)$.   For a
single superconducting grain $\mathcal{L}$ is found by solving the linearized Ginzburg-Landau
equation with the boundary condition corresponding to the zero current flow
at the grain surface. It has the standard form \cite{LV05} at $\epsilon \ll
1 $ but with quantized momenta $q_{k}$. In \ what follows we will need the
analytic continuation of this propagator to the upper half-plane given by
\begin{equation}
{L}^{\text{R}}\left( q_{k},\Omega \right) =-\frac{1}{\nu }\frac{1}{%
\epsilon -i\pi \Omega /8T_{\text{c}}+\xi _{\text{g}}^{2}q_{k}^{2}}.
\label{L}
\end{equation}%
Here $\nu \sim 1/d^{3}\delta $ is the density of states (DoS), and the
quantized momenta are defined by $q_{k}d=2\tan \left( q_{k}d/2\right) $
which gives $q_{0}=0$ and $q_{k}d\approx 2\pi k+\pi $ for $k\geq 1$.
Finally, the shaded triangles in Fig.~\ref{tundia} are the Cooperons
describing the usual `dressing' of the fluctuation propagator due to the
coherent electron scattering from impurities,
\begin{equation}
\quad \lambda \left(  {q}_k,\varepsilon _{1},\varepsilon _{2}\right) =\frac{%
\tau ^{-1}}{|\varepsilon _{1}-\varepsilon _{2}|+\mathcal{D}q_k^{2}}.
\label{Coop}
\end{equation}

Let us first consider the immediate vicinity of $T_{\text{c}}$, region (\ref{3}) where the intergrain coherence length $\xi _{\text{T}}(\epsilon )$ exceeds the grain size.   Then that the granularity is practically irrelevant and the system behaves as the effective amorphous metal with the diffusion coefficient $D_{\text{T} }$. In this region the fluctuation correction $\sigma$  is dominated by the Aslamazov -- Larkin (AL) diagram, Fig.~\ref{tundia}c, and increases in accordance with the standard $3D$ result \cite{LV05}:
\begin{equation}
\sigma _{AL}^{\left( \mathrm{eff}\right) }(\epsilon )\sim \frac{e^{2}}{\xi
_{ \text{T}}}\,\frac{1}{\sqrt{\epsilon }}\,.  \label{totamor}
\end{equation}

On the contrary, in regions (\ref{1}) and  (\ref{2}) the granularity is paramount. We will show that this may result in the appearance of a maximum in the temperature dependence of resistance.
We start our consideration for this region with the  first order fluctuation contribution to $\sigma $, Eq.~(\ref{K}), corresponding to the DoS fluctuating separately in each of
the grains (two such contributions are shown in Fig.~\ref{tundia}a,b). At $\epsilon \ll 1$ the fluctuation propagator, Eq.~(\ref{L})), should be taken at $\Omega=0$. Integrating over the electron momenta and summing
over the fermionic frequencies one finds the following \emph{negative}
contribution:
\begin{equation}
\sigma _{\text{DoS}}(\epsilon )\sim -\,\frac{e^{2}}{d}\,\epsilon _{\text{T}%
}\sum_{k}\frac{1}{\epsilon +\xi _{\text{g}}^{2}q_{k}^{2}}.  \label{DOSgen}
\end{equation}%
In region (\ref{1}), \emph{i.e.}\ for the intragranular $3D$ motion, the
summation in Eq.~(\ref{ALgen}) is reduced to integration, while in the $0D$
region (\ref{2}) only the $q_{0}=0$ term in Eq.~(\ref{DOSgen}) contributes
which gives
\begin{equation}\label{DOSass}
\sigma _{\text{DoS}}(\epsilon )\sim -\,\frac{e^{2}}{d}\epsilon _{\text{T}%
}\times \left\{
\begin{matrix}
\dfrac{1}{\sqrt{\epsilon _{\text{g}}\epsilon }}, & \epsilon _{\text{g}%
}\lesssim \epsilon \lesssim 1 \\[10pt]
\dfrac{1}{\epsilon }, & \epsilon _{\text{T}}\lesssim \epsilon \lesssim
\epsilon _{\text{g}}%
\end{matrix}%
\ \right. .
\end{equation}%

The AL contribution to the correlation function of Eq.~(\ref{K}), diagram \ref{tundia}c \cite{noteLVV3}, is of the second order in $\epsilon_{\text{T}}\equiv \Gamma /T_{\text{c}} $, but it is more
singular in $1/\epsilon $ than the first-order DoS one above. It
is given by
 \begin{align*}
\mathcal{K}_{\text{AL}}\!\left( \omega _{\nu }\right) \!=\!T\!\sum_{\Omega _{n}}\sum_{k,l} \big|B_{kl}\big|^{2}
\mathcal{L} \left( q_{k},\Omega
_{n }\!+\! {\omega_\nu }\right) \mathcal{L} \left( q_{l},\Omega
_{n}\right)  .
\end{align*}%
Here $B_{kl}$ denotes the loop made of four Green's functions (Fig.~\ref%
{tundia}.c) with the Cooperon dressing (\ref{Coop}). The standard
calculation \cite{LV05} gives  $B_{kl}\sim \nu \epsilon_{\text{T}} $. Now one transforms the
sum over $\Omega _{n}$ into a contour integral in a usual way, making
appropriate cuts in the complex plane of the frequency $z$ and thus
constructing the analytical continuation ${\mathrm{i}}\omega _{\nu
}\rightarrow \omega +{\mathrm{i}}0$ which yields in the limit $\omega
\rightarrow 0$ the following expression for the AL contribution into the
fluctuation conductivity:
\begin{align}
\sigma _{\text{AL}}(\epsilon )& \sim \frac{e^{2}}{d}\epsilon _{\text{T}%
}^{2}\!\!\int\limits_{-\infty }^{\infty }\!\!\mathrm{d}\zeta \left( \sum_{k}%
\frac{1}{\left( \epsilon \!+\!\xi _{\text{g}}^{2}q_{k}^{2}\right) ^{2}+\zeta
^{2}}\right) ^{\!2}  \notag \\
& \sim \frac{e^{2}}{d}\epsilon _{\text{T}}^{2}\times \left\{
\begin{array}{ll}
\dfrac{1}{\epsilon _{\text{g}}\epsilon ^{2}}, & \epsilon _{\text{g}}\lesssim
\epsilon \lesssim 1 \\[10pt]
\dfrac{1}{\epsilon ^{3}}, & \epsilon _{\text{T}}\lesssim \epsilon \lesssim
\epsilon _{\text{g}}%
\end{array}%
\ \right. .  \label{ALgen}
\end{align}%

Two types of the Maki -- Thompson (MT) contribution are represented by diagrams in
Fig.~\ref{tundia}d,e. On the face of it, diagram \ref{tundia}d is just a second order contribution from the DoS fluctuations. But this is not
so. The phase coherence of the fluctuation propagators in two grains is
absolutely essential here so that  one may classify this diagram as belonging to the Maki -- Thompson type. Its leading contribution  describes the interference between the
DoS fluctuations in the neighboring grains while only the sub-leading one  contributes to the second order DoS corrections. Technically, it results from the  summation over the anomalous interval where
the fermionic frequencies in the  Cooperons (shaded triangles, each
given by Eq.~(\ref{Coop})) are of the opposite sign. This results in
 the appearance, alongside with the two fluctuation
propagators, of an additional strongly singular factor, cut off by the intragrain dephasing rate $\gamma_\phi = 1/(T_{\text{c}}\tau_\phi ) $:
\begin{equation}
\sigma _{\text{MT}}\sim \frac{e^{2}}{d}\frac{\epsilon _{\text{T}}^{2}}{g_{%
\text{T}}}\sum_{k,l}\frac{1}{ \epsilon \!+\!\xi _{\text{g}%
}^{2}q_{k}^{2} }\frac{1}{ \epsilon \!+\!\xi _{\text{g}%
}^{2}q_{l}^{2} }\frac{1}{\left[ 2\gamma _{\varphi }\! +\!\xi _{\text{g}%
}^{2}q_{k}^{2}+\!\xi _{\text{g}}^{2}q_{l}^{2}\right] ^{3}}\,.  \notag
\end{equation}%
Although this contribution is of the first order in $\Gamma$, its overall factor $\epsilon_{\text{T}}^2/g_{\text{T}}  $ is smaller than that in the AL contribution, which is of the second order in $\Gamma$. The extra factor $1/g_{\text{T}}  \equiv \delta/\Gamma  $ is due to the reduction of the effective phase volume in the former contribution which contains less independent integrations over fast electronic momenta. A
straightforward estimation of the above summation gives
\begin{equation*}
\sigma _{\text{MT}}(\epsilon )\sim \frac{e^{2}}{d}\frac{\epsilon _{\text{T}%
}^{2}}{g_{\text{T}}\gamma _{\varphi }^{3}}\!\times \!\left\{
\begin{array}{ll}
\dfrac{\gamma _{\varphi }}{\epsilon _{\text{g}}\epsilon ^{2}}, &
\epsilon _{\text{g}}\lesssim \gamma _{\varphi }\lesssim \epsilon \lesssim 1
\\[10pt]
\dfrac{1}{\epsilon _{\text{g}}\epsilon },\, & \epsilon
_{\text{g}}\lesssim \epsilon \lesssim \gamma _{\varphi }\lesssim 1 \qquad \quad (12) \\[10pt]
\dfrac{1}{\epsilon ^{2}},\, & \epsilon _{\text{T}
}\lesssim \min\{\epsilon,\gamma_\phi \} \lesssim \epsilon _{\text{g}}
\end{array}%
\right.
\end{equation*}%
The magnitude of this contribution differs from the AL one, Eq.~(\ref{ALgen}), by the factor $g_{\text{T}}\gamma_\phi ^3 $, which may be either small or large. The dephasing rate $\gamma_\phi $ is the sum of
 the escape rate, $\gamma^{{\text{esc}} } _\phi~\sim \Gamma/T_{\text{c}} \equiv \epsilon_{\text{T}}  $, and the  dephasing rate due to the electron-electron interaction, $\gamma^{ee}_\phi =\epsilon_{\text{T}}/(g_{\text{T}}\epsilon_g^{\alpha})    $ with $\alpha=3/2$  in region (\ref{1}) and $\alpha=2$ in region (\ref{2})   \cite{note3}.

The contribution of diagram \ref{tundia}e,
which is a modification  of the standard MT diagram for the
present case, is  smaller  by the factor $\epsilon_{\text{T}} $ than the expression in Eq.~(12).

Let us discuss the results obtained. On the face of it, Eq.~(\ref{ALgen})
contradicts to the well known results for the conductivity \cite{LV05}, $%
\sigma _{AL}^{\left( D\right) }\sim \epsilon ^{\frac{D}{2}-2}$, which are
supposed to be applicable  to any dimensionality $D$.  This apparent discrepancy is due
to the tunneling character of the fluctuation Cooper pair motion between
grains. The Cooper pair tunneling at $T>T_{\text{c}}$  requires two independent electron hops, the probability of each
being proportional to the tunneling rate $\Gamma $. To preserve
the superconducting coherence both should occur within the
Cooper pair life-time $\tau _{\mathrm{GL}}  \sim 1/\epsilon T_{\text{c}} $. Thus the probability $W $ of the
  pair tunneling between two grains $W=\Gamma ^{2 }\tau _{
\mathrm{GL}}\propto 1/\epsilon$ acts as the correction factor to the
expected $0D$ expression $\propto\epsilon ^{-2}$ which leads us to the
result (\ref{ALgen}). In the same spirit one can obtain qualitatively the $3D$
asymptotic of Eq.~(\ref{ALgen}), valid when   $\xi _{g}\ll d$. The
fluctuation conductivity in such a $3D$ grain is the standard $\sigma _{\text{AL}%
}^{\left( 3\right) }\sim \epsilon ^{-1/2}$. Then the  probability $W$ above is reduced due to the fact that only
the pairs in the
skin layer of thickness $\xi _{\text{g}}\left( \epsilon \right) $ near  the grain boundary   can
tunnel, their fraction in the grain being $\xi _{\text{g}}\left( \epsilon
\right) /d$. This immediately leads to the appropriate
asymptotics of Eq.~(\ref{ALgen}).

\begin{figure}[t]\vspace*{2mm}

\includegraphics[width=.5\columnwidth,angle=270]{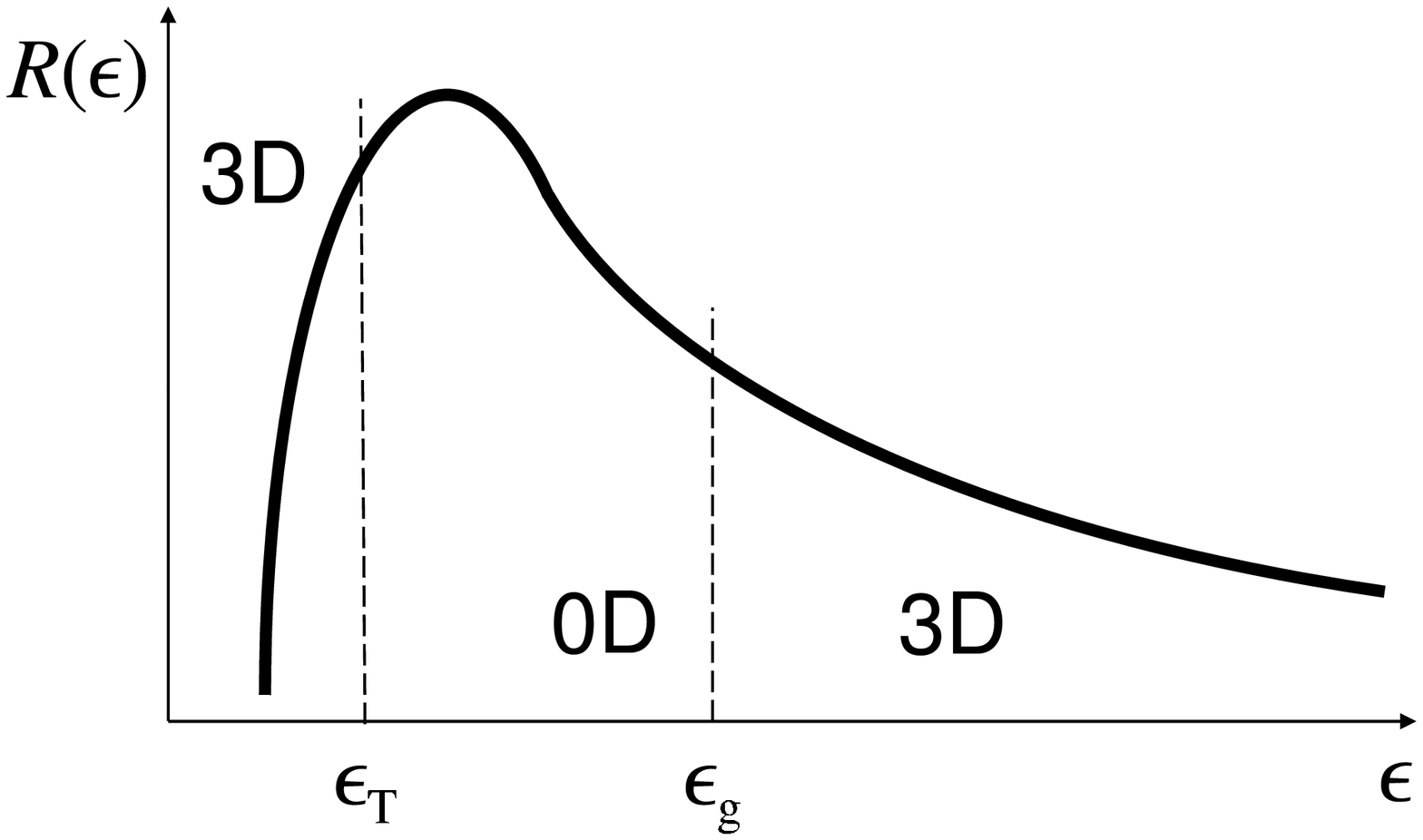}
\caption{Temperature dependence of the granular metal resistance in the
vicinity of superconducting transition for $\epsilon _{\text{g}
}^2\lesssim  \epsilon_{\text{T}} .$ }
\label{resist}
\end{figure}

Finally let us discuss the temperature dependence of the fluctuation contribution to the conductivity given by the sum of Eqs.~(\ref{DOSass})--(12). For $\gamma_\phi ^3 g_{\text{T}}\gg1 $, it is determined by the competition of the negative $\sigma_{\text{DoS}} $ and positive $\sigma_{\text{AL}} $.     At the onset of the fluctuation region, $\epsilon\lesssim 1$, the former dominates since it is proportional to the lowest power of the small tunneling parameter $\epsilon_{\text{T}} $ resulting in the initial increase of the resistance with decrease of $\epsilon\equiv T/T_{\text{c}}-1 $.  With further decrease of $\epsilon$, the AL contribution, more singular in $\epsilon^{-1} $ inevitably wins. The resulting maximum in the  temperature dependence of resistance $R(\epsilon)$ occurs in region (\ref{1}) for $\epsilon_{\text{g}} ^2<\epsilon_{\text{T}} $, or in region (\ref{2}) in the opposite case -- the latter is illustrated in Fig.~\ref{resist}.

In the case of weak dephasing, the MT contribution, Eq.~(12), takes over the AL one, Eq.~(\ref{ALgen}). If dephasing is so small that $ \gamma_\phi ^3 g_{\text{T}}\ll\epsilon_{\text{T}} $, it dominates already at $\epsilon\sim1$ leading to a monotonic decrease in $R(\epsilon)$. For $\epsilon_{\text{T}}\ll\gamma_\phi ^3 g_{\text{T}}\ll1 $ the temperature dependence of resistance  remains qualitatively the same as  in Fig.~\ref{resist}.
 However, the position of the maximum is determined now by $\gamma_\phi $ which makes it sensitive to a weak magnetic field.  Such a field reduces $\tau_\phi $ resulting in the appearance of positive magnetoresistance and a shift of the maximum in $R(\epsilon)$ to lower temperatures. This is in a qualitative agreement with recent experimental measurements \cite{gold07}. Note that the importance of superconducting fluctuations, \textit{e.g.} for the magnetoresistance at low $T$ , is well known \cite{Efetov1}. However, neither the temperature dependence of $R$ near $T_{\text{c}} $  nor the novel MT
mechanism  were considered before.

In conclusion, we have demonstrated that the resistance of the effectively granular  system, characterized by inequalities (\ref{0}), may have a pronounced maximum as $T$ approaches $T_{\text{c}} $.  This maximum is
due to the competition between the fluctuation suppression of the one-electron tunneling between grains with the  enhancement of transport due to
 the coherent charge transfer of the fluctuation Cooper pairs.
 Only at the very edge of the transition  the decay in the
resistance becomes similar to the monotonic one in the amorphous system. Such a distinctive feature together with the possible sensitivity to weak magnetic fields can serve as a
benchmark of the effective granularity of a disordered system.

 We thank Allen Goldman and Igor Aleiner for useful discussions of the results. We are grateful to National Center of Theoretical Sciences of R.O.C.\ for warm hospitality at the final stage of the work. The work was supported by the contract No.\ DE-AC02-06CH11357 of the U.S.\
DoE Office of Science.  I.V.L.\ acknowledges support of the  {EPSRC} (EP/D031109) and
of the NATO collaborative grant. A.A.V.\ acknowledges support of the MIUR
under the project PRIN 2006  and of the
grant RFBR-07-02-12058.


\begin{references}
\bibitem{noteLVV1} {If the intragrain electron motion is ballistic (a
    clean superconductor), then the correlation length $\xi_{\text{g}}$
should be substituted by $\xi_{\text{g}}^{\text{b}}= { v _{\text{F}}
}/{\Delta}$ where   $ v _{\text{F}}$ is the Fermi velocity in the grain. }
\bibitem{noteLVV2} A weak disorder in the grain size or shape and in the
    tunneling amplitude would be irrelevant.
\bibitem{noteLVV3} Other contributions of these type vary in
    different placements of the tunneling amplitudes -- circled
    crosses -- on each fermionic loop. A full account of all
    the AL diagrams is necessary only for an irrelevant
    numerical coefficient. The same is valid for the MT diagrams.
\bibitem{note3}
Here we have expressed $\gamma_\phi=(T_{\text{c}} \tau_\phi )^{-1}  $ via the standard expressions for $\tau_\phi $ in the $3D$ \cite{AAK} and $0D$ \cite{SIA} cases.
\end{references}
\end{document}